\documentclass[twocolumn,showpacs,preprintnumbers,aps,amsmath,amssymb]{revtex4}

\usepackage{graphicx}
\usepackage{dcolumn}
\usepackage{bm}
\usepackage{epsfig}

\begin{document}


\title{Origin of the Kohlrausch exponent}

\author{U. Buchenau}
 \email{buchenau-juelich@t-online.de}
\affiliation{%
J\"ulich Center for Neutron Science, Forschungszentrum J\"ulich\\
Postfach 1913, D--52425 J\"ulich, Federal Republic of Germany
}%

\date{January 15, 2017}

\begin{abstract}
According to a recent numerical finding, the dynamics of a glass former is exclusively due to the forces within the first coordination shell. This implies that the Kohlrausch $\beta$ should be understandable in terms of the effective nearest-neighbor potential.

The present paper proposes a relation for the Kohlrausch $\beta$ based on the Adam-Gibbs conjecture of a flow barrier proportional to the number of atoms or molecules in a cooperatively rearranging region. The conjecture implies that $\beta$ is given by the ratio of the structural entropy increase per particle to the barrier increase per particle. In a recent numerical determination of the structural entropy per particle in Lennard-Jones-like potentials, the relation leads to values between 0.2 and 0.6.
\end{abstract}

\pacs{78.35.+c, 63.50.Lm}
\maketitle

Nearly two centuries ago, Kohlrausch \cite{kohl} observed that the stretched exponential
\begin{equation}\label{k}
	\Phi(t)=\exp(-(t/\tau)^\beta),
\end{equation}
with a Kohlrausch exponent $\beta$ close to 1/2, described the decay of the charge content of a Leyden jar with a glass dielectricum better than a simple single exponential. Much later, it became clear that the stretched exponential relaxation is a general characteristic for the relaxation of the disordered state of matter \cite{ww,bnap}.

Our present understanding of the Kohlrausch relaxation in undercooled liquids is that it results from dynamical heterogeneity \cite{hetero}, i.e. that it must be understood in terms of a sum of many single exponential relaxations occurring in different parts of the sample. But this explanation does not supply a specific value of the Kohlrausch exponent. A stretching exponent $\beta$ requires a marked increase of the total relaxation strength of the local processes with $\tau_r^\beta$ ($\tau_r$ local single-exponential relaxation time).

One has to distinguish between thermally activated local back-and-forth jumps between inherent states (the retardation processes) and the final viscous flow \cite{ferry}, which terminates the local back-and-forth jumps. Mechanical recoverable compliance measurements \cite{plazek-magill,plazek-bero} show that the viscous flow occurs when the retardation response from the local back-and-forth jumps begins to be the same as the elastic response to an applied shear stress. In other words, the Maxwell time $\eta/G$ ($\eta$ viscosity, $G$ infinite frequency shear modulus) is reached when the retardation response is of the same order as the elastic response. 

This conclusion from the recoverable compliance measurements \cite{plazek-magill,plazek-bero} has been recently corroborated by a new interpretation of a large number of dynamical shear data \cite{bu2016} in terms of a pragmatical model for the crossover from retardation processes to the viscous flow.

The last decade has brought a number of important results enlightening the nature of the retardation processes. The exact value of the Kohlrausch exponent $\beta$ of dielectric data  was accurately determined \cite{albena} by measuring the minimum negative slope of the imaginary part $\epsilon''(\omega)$ on the right side of the $\alpha$-peak for 53 glass formers.  The data showed a prevalence for the value $\beta=1/2$, most of the values lying between 0.4 and 0.6. The temperature dependence of these values was small, showing a tendency towards 1/2 with decreasing temperature. 

Further relevant information comes from nonlinear dielectric data \cite{weinstein,brun3,bauer3}. They provide evidence for the proportionality of the flow barrier to the increasing number $N$ of particles in the rearranging core with decreasing temperature (at least in one of the theoretical interpretations \cite{brun3,bauer3}; not in the other \cite{weinstein}). 

The present paper proposes an explanation for the tendency to a Kohlrausch $\beta$ of 1/2 based on this concept. Adding one particle to the rearranging core is supposed to increase the barrier $E_B$ for the rearrangements by $\Delta E_B$. Then $\tau_r$ increases by $\exp(\Delta E_B/k_BT)$.

The same addition increases the number of rearrangement possibilities (i.e. the relaxation strength) by $\exp(S_1/k_B)$, where $S_1$ is the structural entropy per particle. This implies that the Kohlrausch $\beta$ is given by
\begin{equation}
	\beta=\frac{TS_1}{\Delta E_B}.
\end{equation}

\begin{figure}[b]
\hspace{-0cm} \vspace{0cm} \epsfig{file=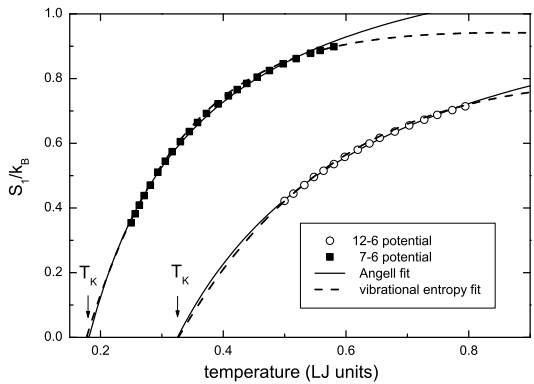,width=7 cm,angle=0} \vspace{0cm} \caption{Numerically determined structural single-particle entropies \cite{kauzmann} in the 12-6 and 7-6 Lennard-Jones potentials. The continuous lines  are fits with the Angell \cite{angell} expression, eq. (\ref{kauz}), the dashed lines fits in terms of structural states with a vibrational entropy proportional to the structural energy (see text).}
\end{figure}

To argue that this value is indeed close to 1/2, we make use of the recent numerical results \cite{kauzmann} in modified Lennard-Jones potentials, in which the exponent of the repulsive part was varied from 12 down to 7, keeping the well depth constant. Fig. 1 shows the calculated structural entropies $S_1$ per particle. The continuous lines are fits in terms of the Angell expression \cite{angell}
\begin{equation}\label{kauz}
	S_1=S_\infty\left(1-\frac{T_K}{T}\right),
\end{equation}
with two parameters, $S_\infty$ and the Kauzmann temperature $T_K$. The dashed lines will be discussed later.

It is seen that the values for $S_1/k_B$ at the low temperature end of the numerically accessible region lie between 0.4 and 0.6.

According to a recent numerical finding \cite{toxvaerd}, the dynamics of a glass former is determined by the forces between the atoms within the first coordination shell. This implies that $\Delta E_B$ should be calculable from the effective nearest-neighbor interatomic potential. To get the values for $\Delta E_B$, let us consider the simplest possible shear rearrangement in close packing in Fig. 2.  

\begin{figure}[b]
\hspace{-0cm} \vspace{0cm} \epsfig{file=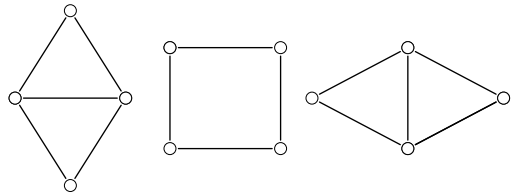,width=7 cm,angle=0} \vspace{0cm} \caption{Elementary shear rearrangement in close packing. To pass from the low energy configuration on the left side to the analogous one on the right side, one has to pass the higher energy square configuration in the middle.}
\end{figure}

The right and left configurations in Fig. 2 have the diagonals 1 and $\sqrt{3}$ in nearest neighbor distances $r_0$, the middle configuration has twice $\sqrt{2}$. For the 12-6 Lennard-Jones potential
\begin{equation}
	V(x)=4\left(\frac{1}{x^{12}}-\frac{1}{x^6}\right)
\end{equation}
with $r_0=2^{1/6}$, this implies a contribution to $\Delta E_B$ of 0.5493 Lennard-Jones units from the interatomic potential, for the 7-6 potential
\begin{equation}
	v(x)=4\left(\frac{2^{1/6}3}{x^7}-\frac{7}{2x^6}\right),
\end{equation}
one calculates 0.4416 Lennard-Jones units. 

With these two contributions, one estimates $\beta=0.4..0.62$ between $T=0.5$ and $T=0.6$ for the 12-6 potential, $\beta=0.2$ to $\beta=0.5$ for the 7-6 potential in the temperature interval between 0.25 and 0.35.

The estimate is crude, for three reasons. The first is that the middle configuration of Fig. 2 is always present, even in an fcc crystal, which has one octahedral hole per atom. The second is that one should add the temperature-dependent forces which keep the atoms at their position, an influence which increases $\Delta E_B$. The third is that one does not only have an energetic contribution to $\Delta E_B$, but also (as will be seen below) an entropic one, an influence which decreases $\Delta E_B$. 

Note that one gets much closer to the Kauzmann temperature (to 1.38 $T_K$) in the 7-6 potential than in the 12-6 potential (1.54 $T_K$). Since the Kauzmann temperature is usually close to the Vogel-Fulcher temperature \cite{angell}, where the viscosity diverges, this implies that the 7-6 potential is the more fragile case. This is strange, because it is decidedly the more harmonic potential of the two. On the other hand, the finding of lower $\beta$ values in this more fragile case is consistent with experiment \cite{bnap}.

Real experiments are usually done at constant pressure, with a pronounced thermal expansion. In that case, $\Delta E_B$ should also increase with increasing temperature, because the nearest neighbor distance increases. Thus $\beta$ could also decrease with temperature at constant pressure; it depends which influence is strongest. 

The argument shows that the anharmonic softening of the shear modulus does not necessarily imply a softening of the flow barrier \cite{nemilov,dyre}. In fact, nonlinear dielectric measurements \cite{bauer3} show a proportionality of the flow barrier to the number of molecules in a cooperatively rearranging region in several substances, with no visible influence of the anharmonicity.

The comparison to real experiments in molecular glasses \cite{angell}, where eq. (\ref{kauz}) was first proposed, shows much larger values of $S_\infty$. In the two Lennard-Jones potentials, one finds $S_\infty/k_B=1.218$ for the 12-6 potential and $S_\infty/k_B=1.331$ for the 7-6 potential. In the molecular glasses \cite{angell}, one finds values between 8 and 17.

Naturally, molecules like salol and orthoterphenyl have much more degrees of freedom than a single atom. But it is still surprising to see a factor of ten. It implies that not only the molecular orientation, but also inner degrees of freedom participate in the creation of structural entropy. One can obviously build more structures by deforming the molecule itself.

On the other hand, deforming and reorienting a molecule does not contribute to the shear flow. For the flow of a molecular glass former, one has to fall back on the three translational degrees of freedom, which enable shear transformations like the one in Fig. 2. Therefore it seems reasonable to assume that only the tenth of the structural entropy due to the translational degrees of freedom has to be taken into account to understand the viscous flow. Thus one expects (and finds) similar Kohlrausch-$\beta$-values in molecular glasses as in the Lennard-Jones cases.

Finally, let us try to replace the notoriously unexplainable Angell expression, eq. (\ref{kauz}), by a more physical picture. The equation tells us that one has to reckon with a finite structural entropy $S_1$. However, if one tries to describe the curves in Fig. 1 in terms of a gaussian in structural energy containing about four states per atom, one fails; it is possible to get the right Kauzmann temperature by adapting the width of the gaussian, but the curvature that one gets is too small to reproduce the measured data.

One gets better agreement if one postulates a larger vibrational entropy $S_{vib}$ for the states with a higher structural energy $E$, say with
\begin{equation}
	S_{vib}=\frac{E}{E_M}S_0
\end{equation}
for states between $E=0$ and $E=E_M$, with the constant density of states $\exp(S_1/k_B)/E_M$ between these two values.
The dashed lines in Fig. 1 show fits, for the 12-6 potential with $E_M=2.63$ Lennard-Jones units, $S_1/k_B=0.96$ and $S_0/k_B=0.6$. The 7-6 potential differs only in the value for $E_M$, 1.443 Lennard-Jones units. Obviously, $E_M$ describes the influence of the specific potential, $S_1$ and $S_0$ characterize a rather general packing entropy.

This description has the advantage to provide a physical picture of the packing entropy, with a reduced number of only 2.6 states per atom, but with a direct understanding of the strong vibrational softening of glass formers with increasing temperature. The maximum vibrational entropy difference of 0.6 $k_B$ is larger than the typical fcc-bcc entropy difference of about 0.15 $k_B$, but smaller than the large vibrational entropy difference 0.9 $k_B$ between white and gray tin \cite{seitz}, with gray tin in the rather rigid diamond structure.

Further evidence for the validity of the description is the surprising proportionality of the logarithm of the viscosity to the inverse of the difference of the mean square displacements of crystal and liquid in selenium \cite{zorn}, orthoterphenyl and glycerol \cite{zorn1}. The description implies a proportionality of this difference to the structural entropy, which in turn is inversely proportional to the logarithm of the viscosity according to the Adam-Gibbs conjecture.

\end{document}